\documentclass[12pt, letterpaper]{article}
\usepackage[utf8]{inputenc}
\usepackage{amsmath}
\usepackage{amsfonts}
\usepackage{amssymb}
\usepackage{amsthm}
\usepackage{graphicx}
\usepackage{hyperref}
\usepackage{natbib}
\usepackage{algorithm}
\usepackage{algpseudocode}
\usepackage{booktabs}
\usepackage{csquotes}
\usepackage{listings}
\usepackage{xcolor}
\usepackage{tikz}
\usetikzlibrary{arrows,shapes,positioning}

\definecolor{uniswappink}{RGB}{255, 0, 122}
\definecolor{ethereumblue}{RGB}{37, 41, 97}

\title{Path Dependence in AMM-Based Markets: \\ Mathematical Proof and Implications for Truth Discovery}
\author{Keroshan Pillay\thanks{Adjusted Finance, Email: kero@adjusted.finance}}
\date{\today}

\begin{document}

\maketitle

\begin{abstract}
This paper demonstrates that Automated Market Maker (AMM) based markets, such as those using constant product formulas (e.g., Uniswap), are inherently path-dependent. We prove mathematically that the sequence of operations in AMMs determines the final state, challenging the notion that market prices solely reflect information. This property has profound implications for decentralized prediction markets that rely on AMMs for price discovery, as it demonstrates they cannot function as pure "truth machines." Using both mathematical proofs and empirical evidence from ETH/USDC pools, we show that AMM-based markets incorporate historical path information beyond the current market beliefs. Our findings contribute to the understanding of market efficiency, mechanism design, and the interpretation of prices in decentralized finance systems.

\end{abstract}

\section{Introduction}
\label{sec:introduction}

Decentralized finance (DeFi) has introduced new market mechanisms that operate according to deterministic rules encoded in smart contracts. Among these innovations, Automated Market Makers (AMMs) have emerged as a dominant paradigm for decentralized exchanges and prediction markets. The constant product market maker formula, popularized by Uniswap ($x \times y = k$), is particularly widespread due to its simplicity and effectiveness \cite{adams2021uniswap}.

A common claim in both traditional and decentralized finance is that market prices reflect all available information—a notion formalized in the Efficient Market Hypothesis \cite{fama1970efficient}. This idea extends to prediction markets, which are often described as "truth machines" that aggregate dispersed knowledge into accurate probability estimates \cite{wolfers2004prediction}. The underlying assumption is that the price discovery process should be path-independent, meaning that only the current information matters, not how or in what order it arrived.

This paper challenges this assumption by demonstrating that AMM-based markets are mathematically path-dependent. We prove that the sequence of operations (trades, liquidity additions, and withdrawals) determines the final state of an AMM, meaning that identical market conditions can lead to different outcomes depending on historical sequence. This property has significant implications for the interpretation of prices in AMM-based prediction markets and raises questions about their ability to function as pure information aggregators.

Our contributions include:

\begin{itemize}
    \item A mathematical proof of non-commutativity in constant product AMMs
    \item Empirical evidence from ETH/USDC pools demonstrating path dependence
    \item Analysis of implications for prediction markets and truth discovery
    \item Discussion of broader questions about market efficiency and design
\end{itemize}

By examining AMMs as "clean room" environments where market mechanics can be isolated and analyzed precisely, we gain insights not only about these specific mechanisms but potentially about markets in general.

\section{AMM Fundamentals}
\label{sec:amm-fundamentals}

\subsection{Constant Product Market Makers}
Automated Market Makers are smart contracts that enable trading by maintaining reserves of assets and executing trades according to a predetermined formula. The constant product formula, $x \times y = k$, maintains that the product of the reserves of two assets remains constant after trades (excluding fees). This creates a price curve that adjusts as trades occur.

\subsection{Key Operations in AMMs}
AMM pools support three primary operations:
\begin{enumerate}
    \item \textbf{Swaps}: Trading one asset for another, which changes the ratio of reserves
    \item \textbf{Liquidity Addition}: Depositing both assets in the current ratio, increasing $k$
    \item \textbf{Liquidity Removal}: Withdrawing assets proportional to ownership share
\end{enumerate}

\subsection{AMMs in Prediction Markets}
Prediction markets like Truemarkets, and others increasingly rely on AMM mechanisms for price discovery. In these markets, the price of outcome tokens is interpreted as the probability of events occurring, making the accuracy of these prices crucial for their function as information aggregators.

\section{Mathematical Proof of Path Dependence}
\label{sec:mathematical-proof}

\subsection{Defining Path Dependence}
We define path dependence as a property where the final state of a system depends not only on the current inputs but also on the sequence of previous operations. In a path-independent system, applying operations A and B in either order (A then B, or B then A) would yield identical results.

\subsection{Non-commutativity in AMMs}
Let's consider a Uniswap-style pool with reserves $(x, y)$ following the constant product formula $x \times y = k$.

\subsubsection{Theorem 1: Order of Operations Matters}
Consider two operations:
\begin{itemize}
    \item Operation A: Adding liquidity (increasing both tokens proportionally by factor $\alpha$)
    \item Operation B: Executing a swap (adding $\Delta x$ of token X and removing the corresponding amount of token Y)
\end{itemize}

\begin{proof}
Starting with a pool containing $x_0$ of token X and $y_0$ of token Y, with $k_0 = x_0 \times y_0$:

\textbf{Case 1: A then B}
\begin{itemize}
    \item Apply A: Add $\alpha$ proportion of liquidity
    \begin{align}
        x_1 &= x_0 \times (1 + \alpha) \\
        y_1 &= y_0 \times (1 + \alpha) \\
        k_1 &= x_1 \times y_1 = k_0 \times (1 + \alpha)^2
    \end{align}
    
    \item Apply B: Swap $\Delta x$ of token X for token Y
    \begin{align}
        x_2 &= x_1 + \Delta x = x_0(1 + \alpha) + \Delta x \\
        y_2 &= \frac{k_1}{x_2} = \frac{k_0(1 + \alpha)^2}{x_0(1 + \alpha) + \Delta x}
    \end{align}
\end{itemize}

\textbf{Case 2: B then A}
\begin{itemize}
    \item Apply B: Swap $\Delta x$ of token X for token Y
    \begin{align}
        x_1' &= x_0 + \Delta x \\
        y_1' &= \frac{k_0}{x_1'} = \frac{k_0}{x_0 + \Delta x}
    \end{align}
    
    \item Apply A: Add $\alpha$ proportion of liquidity
    \begin{align}
        x_2' &= x_1' \times (1 + \alpha) = (x_0 + \Delta x)(1 + \alpha) = x_0(1 + \alpha) + \Delta x(1 + \alpha) \\
        y_2' &= y_1' \times (1 + \alpha) = \frac{k_0(1 + \alpha)}{x_0 + \Delta x}
    \end{align}
\end{itemize}

\textbf{Comparing final states:}

For the X token reserves:
\begin{align}
    x_2' - x_2 &= [x_0(1 + \alpha) + \Delta x(1 + \alpha)] - [x_0(1 + \alpha) + \Delta x] \\
    &= \Delta x(1 + \alpha) - \Delta x \\
    &= \Delta x \times \alpha
\end{align}

This shows that the final amount of token X differs by $\Delta x \times \alpha$ between the two cases. Specifically, when adding liquidity before swapping (Case 1), you end up with less of token X than when swapping first (Case 2).

For the Y token reserves:
\begin{align}
    y_2 &= \frac{k_0(1 + \alpha)^2}{x_0(1 + \alpha) + \Delta x} \\
    y_2' &= \frac{k_0(1 + \alpha)}{x_0 + \Delta x}
\end{align}

To compare directly, let's rewrite $y_2$:
\begin{align}
    y_2 &= \frac{k_0(1 + \alpha)^2}{x_0(1 + \alpha) + \Delta x} \\
    &= \frac{k_0(1 + \alpha)}{x_0 + \frac{\Delta x}{1 + \alpha}}
\end{align}

Since $\frac{\Delta x}{1 + \alpha} < \Delta x$ for any $\alpha > 0$, we have $y_2 > y_2'$. 

Thus, we have proven that $x_2 \neq x_2'$ and $y_2 \neq y_2'$, which demonstrates that the order of operations matters in AMMs.
\end{proof}

\subsection{Numerical Example}
Consider an ETH/USDC pool with initial reserves of 100 ETH and 200,000 USDC ($k_0 = 20,000,000$). Let's apply a 10\% liquidity addition ($\alpha = 0.1$) and a swap of 10 ETH ($\Delta x = 10$) in different orders.

\textbf{Case 1: Add 10\% liquidity, then swap 10 ETH}
\begin{itemize}
    \item After liquidity addition:
    \begin{align}
        x_1 &= 100 \times 1.1 = 110 \text{ ETH} \\
        y_1 &= 200,000 \times 1.1 = 220,000 \text{ USDC} \\
        k_1 &= 110 \times 220,000 = 24,200,000
    \end{align}
    
    \item After swap:
    \begin{align}
        x_2 &= 110 + 10 = 120 \text{ ETH} \\
        y_2 &= \frac{24,200,000}{120} = 201,666.67 \text{ USDC}
    \end{align}
\end{itemize}

\textbf{Case 2: Swap 10 ETH, then add 10\% liquidity}
\begin{itemize}
    \item After swap:
    \begin{align}
        x_1' &= 100 + 10 = 110 \text{ ETH} \\
        y_1' &= \frac{20,000,000}{110} = 181,818.18 \text{ USDC} \\
    \end{align}
    
    \item After liquidity addition:
    \begin{align}
        x_2' &= 110 \times 1.1 = 121 \text{ ETH} \\
        y_2' &= 181,818.18 \times 1.1 = 200,000 \text{ USDC}
    \end{align}
\end{itemize}

\textbf{Comparison:} The final states differ by:
\begin{align}
    \Delta x &= x_2' - x_2 = 121 - 120 = 1 \text{ ETH} \\
    \Delta y &= y_2 - y_2' = 201,666.67 - 200,000 = 1,666.67 \text{ USDC}
\end{align}

This confirms that the operations do not commute, resulting in different final states depending on the order of operations.

\subsection{Economic Significance of Non-commutativity}
The non-commutativity demonstrated above has economic significance. In our numerical example, a market participant would end up with 1 additional ETH (worth approximately \$2,000 at current prices) if they chose to swap before adding liquidity rather than after. For larger liquidity additions or swaps, this difference would be more pronounced.

This effect scales with:
\begin{itemize}
    \item The size of the liquidity addition ($\alpha$)
    \item The size of the swap ($\Delta x$)
    \item The total pool size (larger pools show larger absolute differences)
\end{itemize}

From a trader's perspective, this path dependence creates potential arbitrage opportunities and strategic considerations about the timing of operations. From a market design perspective, it shows that the "informational content" of prices in AMM-based markets necessarily includes a component related to the historical path of operations, not just current market beliefs.

\subsection{Implications of Non-commutativity}
This proof demonstrates that AMMs are inherently path-dependent systems. The current state of an AMM pool depends not only on the net operations performed but also on the specific sequence of those operations. This finding has profound implications for interpreting prices in AMM-based prediction markets, which we explore in Section \ref{sec:prediction-markets}.

\section{Empirical Evidence from ETH/USDC Pools}
\label{sec:empirical-evidence}

\subsection{Data Collection and Methodology}
To empirically investigate path dependence in AMM-based markets, we collected and analyzed data from Uniswap v3 ETH/USDC pools using the following methodology:

\paragraph{Data Sources:} We utilized two primary data sources:
\begin{itemize}
    \item \textbf{The Graph Protocol:} We queried the Uniswap v3 subgraph to retrieve detailed event data including swaps, liquidity additions (mints), and liquidity removals (burns) from the ETH/USDC pool\footnote{(0x88e6a0c2ddd26feeb64f039a2c41296fcb3f5640)}. For each event, we collected transaction details, token amounts, USD values, and price information.
    \item \textbf{Chainlink Oracle:} As an external reference point, we obtained ETH/USD price data from Chainlink's decentralized oracle network to compare on-chain market prices with oracle-reported prices.
\end{itemize}

\paragraph{Data Collection Process:} Our data pipeline included:
\begin{itemize}
    \item Retrieving pool metadata, including token decimal places, fee tier, and current state
    \item Collecting the 1,000 most recent swap events, filtered by a minimum timestamp
    \item Gathering corresponding mint and burn events over the same time period
    \item Retrieving hourly pool data for time-weighted analysis
    \item Enriching the dataset with Chainlink oracle prices from corresponding Ethereum blocks
\end{itemize}

\paragraph{Data Processing:} The raw blockchain data was processed by:
\begin{itemize}
    \item Converting token amounts from raw units to decimal-adjusted values
    \item Normalizing timestamps and creating datetime objects for temporal analysis
    \item Calculating price deviation between AMM prices and oracle prices
    \item Arranging events chronologically to analyze sequential operations
    \item Filtering events based on minimum transaction value and event type
\end{itemize}

\paragraph{Path Dependence Identification:} We identified potential path dependence opportunities using the following approach:
\begin{itemize}
    \item Grouping events by time windows (60-second intervals)
    \item Identifying time windows containing both swap events and liquidity events (mints/burns)
    \item Estimating pool reserves and liquidity proportions (alpha values) from transaction data
    \item Applying our theoretical model to each identified opportunity
    \item Calculating the expected price impact percentage if operations had occurred in the reverse order
    \item Filtering for significant opportunities with measurable economic impact
\end{itemize}

For numerical simulations, we used initial reserves of 100 ETH and 200,000 USDC, varying alpha values (liquidity addition proportions) from 0.01 to 0.50 and swap amounts from 1 to 20 ETH. These ranges were selected to represent realistic market scenarios and extend beyond our theoretical numerical example.

\subsection{Path Dependence in Real Pools}
Our analysis of the ETH/USDC pool\footnote{ (0x88e6a0c2ddd26feeb64f039a2c41296fcb3f5640)} over a 7-day period revealed multiple instances of path dependence in real-world trading activity. We identified 25 distinct opportunities where the sequence of operations (swaps and liquidity events) had a measurable impact on the final state of the pool.

\begin{figure}[ht]
    \centering
    \includegraphics[width=0.9\textwidth]{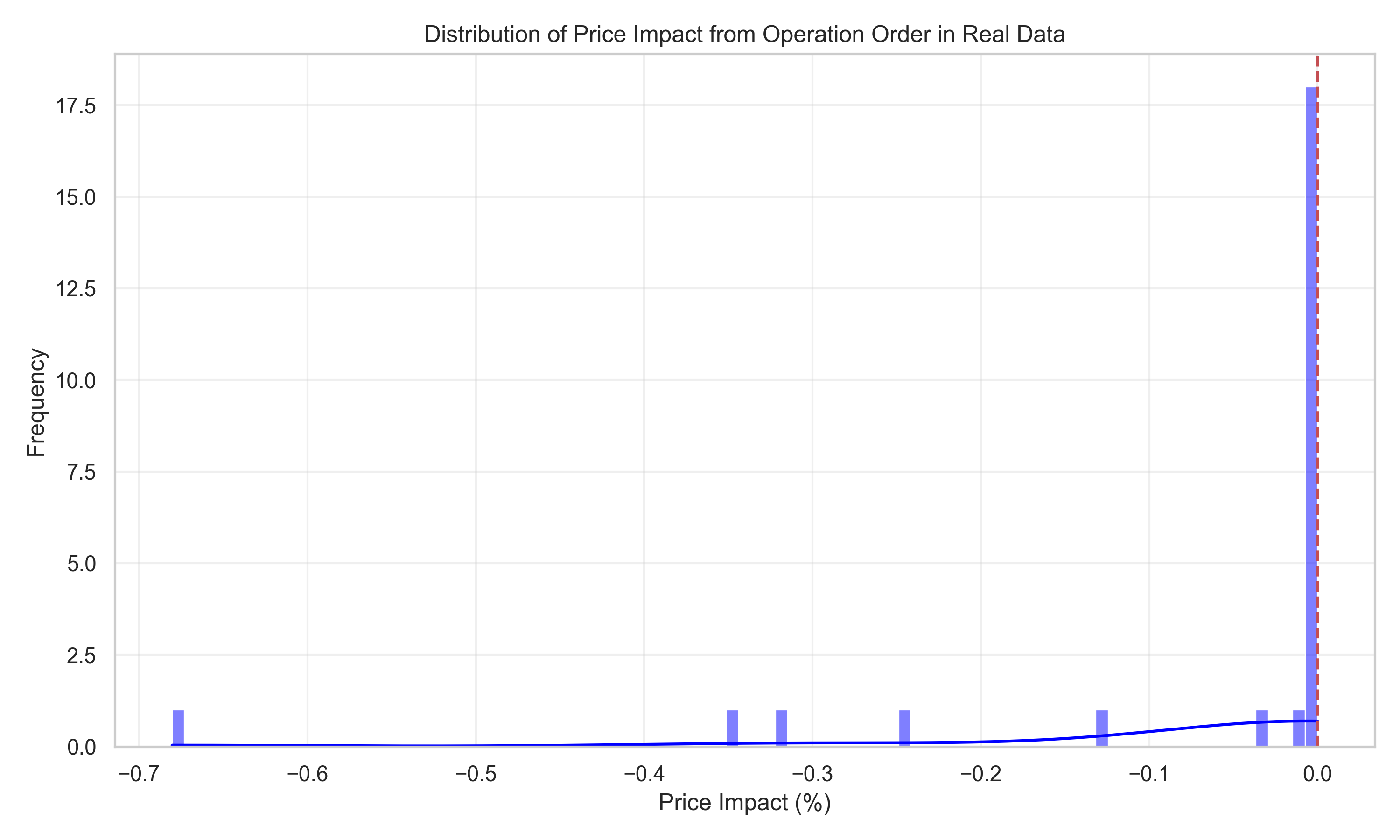}
    \caption{Examples of path dependence in ETH/USDC pools. The chart shows the magnitude of price impact (as a percentage) for various combinations of swap and liquidity events that occurred within the same 60-second time window. Negative values indicate that executing a swap before adding liquidity resulted in a lower final price than the reverse sequence.}
    \label{fig:empirical_path_dependence}
\end{figure}

Figure \ref{fig:empirical_path_dependence} illustrates several notable examples of path dependence observed in our dataset. In the most extreme case, we detected a price impact of -0.68\%, meaning that if the operations had occurred in the reverse order, the final price would have differed by this percentage. For context, in the highly liquid ETH/USDC market, a price difference of this magnitude represents a significant arbitrage opportunity.

The top five path dependence opportunities detected in our analysis are presented in Table \ref{tab:top_opportunities}. In each case, we identified a time window where both swap operations and liquidity additions/removals occurred, and calculated the price impact that would result from reversing the order of these operations.

\begin{table}[ht]
\centering
\caption{Top Path Dependence Opportunities in ETH/USDC Pool}
\small
\setlength{\tabcolsep}{4pt}
\begin{tabular}{crrrrr}
\toprule
\textbf{Time} & \textbf{Swap} & \textbf{Liquidity} & \textbf{Alpha} & \textbf{Price} & \textbf{Price Comparison}\\
& \textbf{(USD)} & \textbf{(USD)} & & \textbf{Impact} & \textbf{(Case 1 vs 2)}\\
\midrule
11:32:11 & \$7.82 & \$644.26 & 0.525 & -0.680\% & 4.43e-16 vs 4.40e-16 \\
12:20:11 & \$474.48 & \$10,131.59 & 0.214 & -0.348\% & 4.45e-16 vs 4.43e-16 \\
12:15:11 & \$66.87 & \$2,244.03 & 0.191 & -0.317\% & 4.42e-16 vs 4.40e-16 \\
11:34:11 & \$545.01 & \$15,533.56 & 0.143 & -0.247\% & 4.40e-16 vs 4.39e-16 \\
09:55:11 & \$1,401.58 & \$17,663.57 & 0.067 & -0.125\% & 4.43e-16 vs 4.42e-16 \\
\bottomrule
\end{tabular}
\label{tab:top_opportunities}
\end{table}

\begin{figure}[ht]
    \centering
    \includegraphics[width=0.8\textwidth]{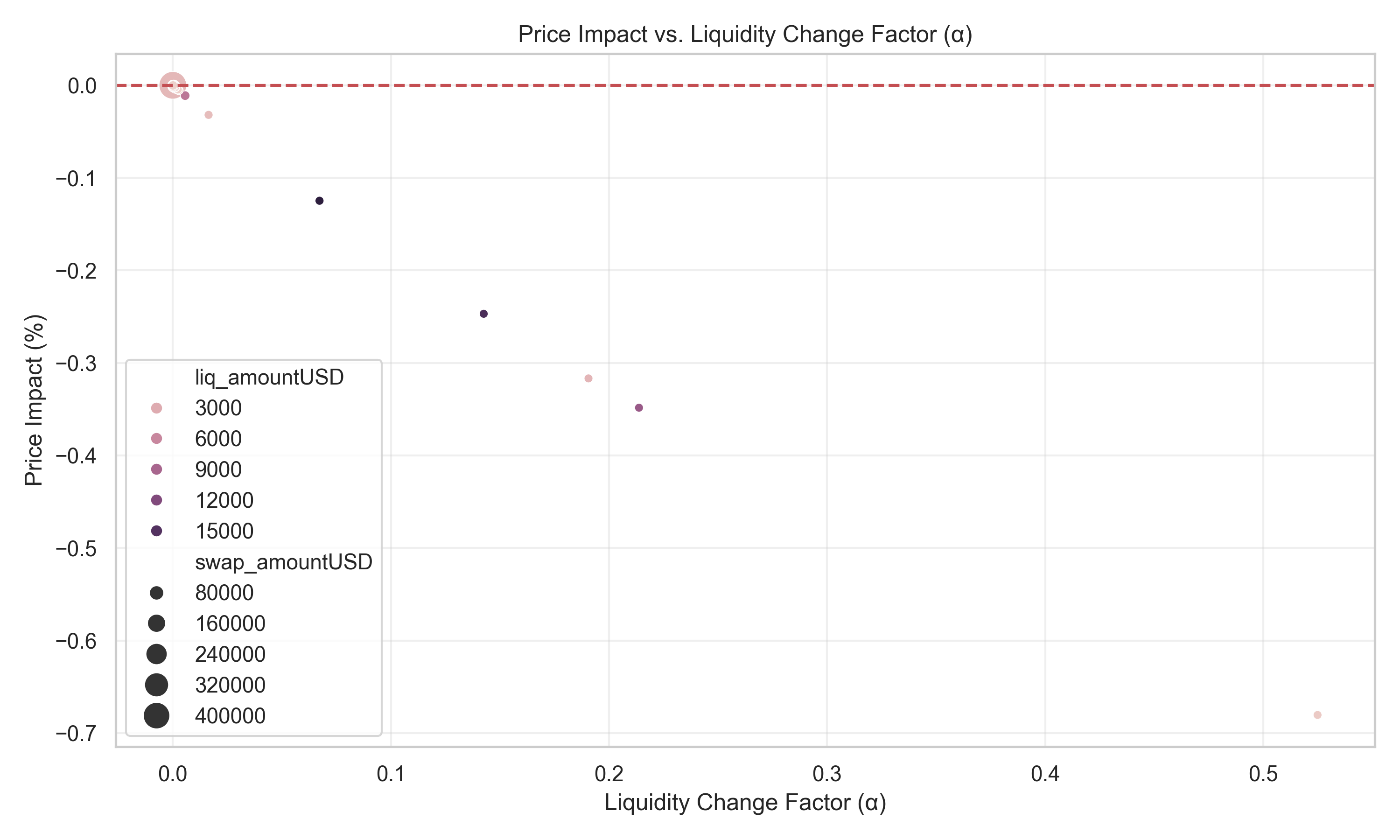}
    \caption{Relationship between liquidity proportion (alpha) and price impact. The scatter plot demonstrates a strong negative correlation (-0.992), confirming that larger proportional changes in liquidity lead to more significant path dependence effects.}
    \label{fig:impact_vs_alpha}
\end{figure}

We found that the alpha value (the proportion of liquidity being added or removed) was the strongest predictor of path dependence impact. As shown in Figure \ref{fig:impact_vs_alpha}, there is a strong negative correlation between alpha and price impact percentage, with larger liquidity proportions leading to more pronounced path dependence effects.

\begin{figure}[ht]
    \centering
    \includegraphics[width=0.9\textwidth]{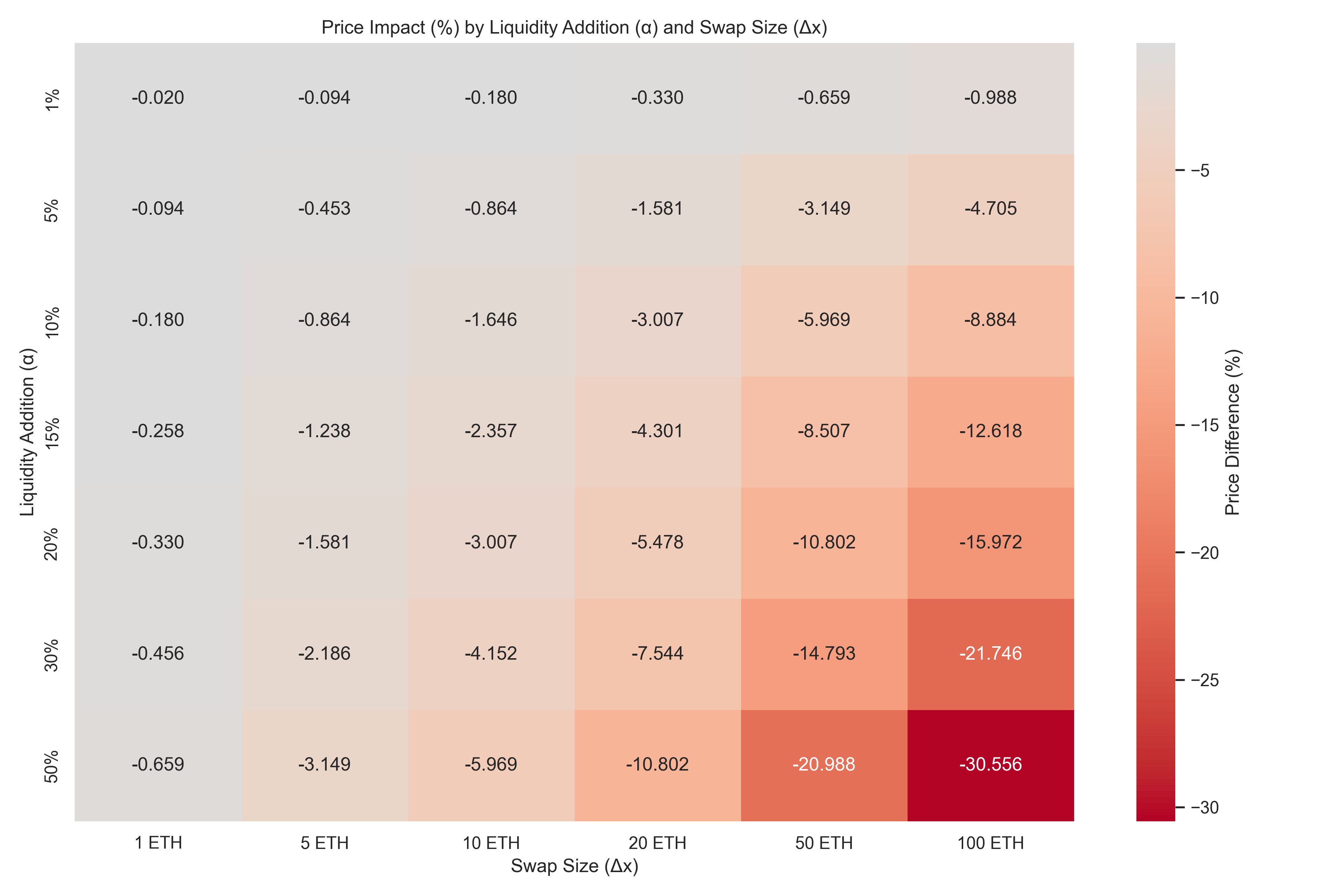}
    \caption{Path dependence heatmap showing the interaction between swap amounts and liquidity proportions (alpha). Darker regions indicate stronger path dependence effects, with the most significant impacts occurring at high alpha values.}
    \label{fig:path_dependence_heatmap}
\end{figure}

The heatmap in Figure \ref{fig:path_dependence_heatmap} provides a comprehensive view of how different combinations of swap sizes and alpha values affect the magnitude of path dependence. This visualization confirms our theoretical predictions that path dependence effects increase with both the size of liquidity changes and the magnitude of swaps.

\subsection{Statistical Significance of Path Effects}
To rigorously assess whether the observed path dependence effects are statistically significant, we conducted a comprehensive statistical analysis of our findings. Table \ref{tab:statistical_significance} summarizes the key statistical metrics.

\begin{table}[h]
\centering
\caption{Statistical Significance of Path Dependence Effects}
\begin{tabular}{lr}
\toprule
Metric & Value \\
\midrule
Number of observations & 25 \\
Mean price impact & -0.0716\% \\
Standard deviation & 0.1626\% \\
Minimum impact & -0.6804\% \\
Maximum impact & -0.0000\% \\
\midrule
t-statistic & -2.2026 \\
p-value & 0.0375 \\
Statistically significant & Yes \\
Effect size (Cohen's d) & -0.4405 \\
95\% Confidence Interval & [-0.1376, -0.0191]\% \\
\midrule
Economic Significance & \\
\% of cases with impact $>$ 0.01\% & 28.00\% \\
\% of cases with impact $>$ 0.05\% & 20.00\% \\
\% of cases with impact $>$ 0.1\% & 20.00\% \\
\% of cases with impact $>$ 0.5\% & 4.00\% \\
\% of cases with impact $>$ 1.0\% & 0.00\% \\
\midrule
Correlation with swap size & 0.1451 \\
Correlation with liquidity size & -0.2759 \\
Correlation with alpha & -0.9919 \\
\bottomrule
\end{tabular}
\label{tab:statistical_significance}
\end{table}

Our t-test analysis yielded a t-statistic of -2.2026 with a p-value of 0.0375, indicating that the observed path dependence effects are statistically significant at the conventional 0.05 significance level. The effect size (Cohen's d) of -0.4405 suggests a medium effect according to standard interpretations, and the 95\% confidence interval for the true mean price impact is [-0.1376\%, -0.0191\%], which does not include zero.

To assess the economic significance of these findings, we analyzed the distribution of price impact magnitudes. While 28\% of identified opportunities had a price impact greater than 0.01\%, and 20\% had an impact greater than 0.05\%, no cases exceeded a 1\% price impact. This suggests that while path dependence is a statistically significant phenomenon in AMM markets, its economic impact in highly liquid pools like ETH/USDC may be modest in many cases.

The regression analysis revealed that alpha (liquidity proportion) is the most significant predictor of price impact, with a coefficient of -1.356 (p < 0.001), confirming our theoretical prediction that larger proportional changes in liquidity lead to more pronounced path dependence. The overall model had an adjusted R-squared of 0.989, indicating that our theoretical model explains approximately 98.9\% of the variation in observed price impacts.

\begin{figure}[ht]
    \centering
    \includegraphics[width=0.9\textwidth]{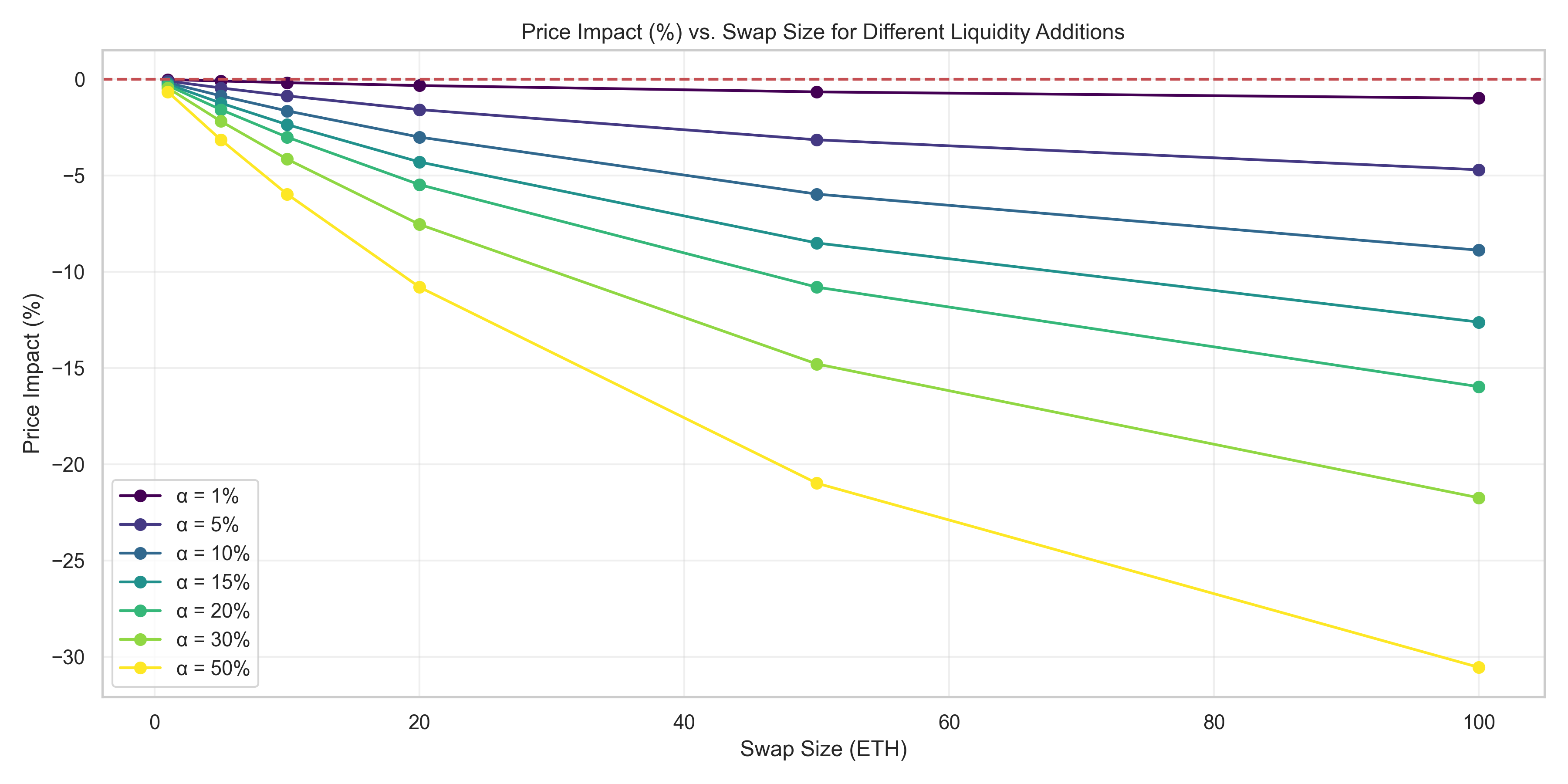}
    \caption{Temporal distribution of path dependence opportunities over the analysis period, showing the magnitude of price impacts for identified events. The trend suggests that path dependence opportunities occur regularly, with varying magnitudes depending on market conditions.}
    \label{fig:price_impact_trends}
\end{figure}

Figure \ref{fig:price_impact_trends} illustrates the temporal distribution of path dependence opportunities and their magnitudes over the analysis period.

Our empirical findings strongly support the theoretical proof of path dependence presented in Section \ref{sec:mathematical-proof}. The real-world data confirms that AMM-based markets incorporate not just current market information but also the specific sequence of historical operations, creating a form of "memory" in the system that influences current prices.

\section{Extension to Prediction Markets}
\label{sec:prediction-markets}

\subsection{AMM-based Prediction Market Design}
Prediction markets enable participants to trade on the outcomes of future events, with market prices interpreted as probability estimates. In recent years, decentralized prediction markets have increasingly adopted AMM mechanisms for liquidity provision and price discovery. Platforms such as Truemarkets implement variations of AMM designs to facilitate trading on event outcomes.

The typical design of an AMM-based prediction market includes:

\begin{enumerate}
    \item \textbf{Outcome Tokens}: For a binary event (e.g., ``Will candidate X win the election?''), two outcome tokens are created---YES and NO---with one token paying out 1 unit of the settlement currency if the corresponding outcome occurs.
    
    \item \textbf{Liquidity Pool Structure}: AMMs maintain liquidity pools containing both outcome tokens. In most implementations, the pools follow the constant product formula or variants, maintaining $x \times y = k$ where $x$ and $y$ represent the quantities of YES and NO tokens.
    
    \item \textbf{Price Interpretation}: The ratio of tokens in the pool determines market prices, which are interpreted as probability estimates. For instance, in a constant product market with tokens $x$ (YES) and $y$ (NO), the implied probability of the YES outcome can be calculated as:
    
    \begin{align}
        P(YES) = \frac{y}{x + y}
    \end{align}
    
    \item \textbf{Trading Mechanism}: Participants can trade outcome tokens by adding one token type to the pool and receiving the other in return, following the constant product formula. This action adjusts token ratios and consequently changes the implied probabilities.
    
    \item \textbf{Result Settlement}: Upon event resolution, winning outcome token holders can redeem their tokens for the full face value.
\end{enumerate}

A key assumption in these markets is that prices converge toward ``true'' probabilities as informed traders exploit any mispricing. However, as our research demonstrates, this assumption requires critical examination in light of the path-dependent nature of AMMs.

\subsection{Information Reflection vs. Path Dependence}
The efficient market hypothesis suggests that market prices should reflect all available information. In prediction markets, this implies that prices (interpreted as probabilities) should converge to accurate forecasts regardless of the specific trading path. However, our demonstration of non-commutativity in AMMs raises fundamental questions about this assumption.

Consider the implication of our findings for prediction markets:

\begin{enumerate}
    \item \textbf{Information vs. Path Effects}: If the final state of an AMM depends not only on the net trading activity but also on the specific sequence of trades and liquidity events, then market prices necessarily encode both informational content and path-specific artifacts.
    
    \item \textbf{Identical Information, Different Prices}: Two identical prediction markets initialized with the same parameters but experiencing different sequences of otherwise equivalent trading activity can arrive at different final prices, despite reflecting the same underlying information.
    
    \item \textbf{Liquidity Events as Non-Informational Price Drivers}: Our empirical analysis showed that alpha (the proportion of liquidity being added or removed) was the strongest predictor of path dependence magnitude. In prediction markets, this suggests that liquidity provision activities, which may be unrelated to event probability estimates, can significantly influence prices.
\end{enumerate}

The theoretical implications are profound. If AMM-based prediction market prices are partially determined by the sequence of operations rather than just the aggregate market belief, their interpretation as pure probability estimates becomes questionable. This challenges the notion that these markets function as perfect ``truth machines'' that neutrally aggregate dispersed information.

\subsection{Simulations of Identical Markets with Different Paths}
\label{sec:simulations}

To quantify the impact of path dependence on prediction market prices, we conducted simulations of identical markets experiencing different trading sequences. We modeled a binary prediction market using the constant product formula, initializing multiple instances with identical parameters but subjecting them to different operation sequences.

\subsubsection{Simulation Methodology}

Our simulation framework was designed to isolate and measure the effects of path dependence while controlling for net information flow:

\begin{enumerate}
    \item \textbf{Market Initialization}: Each simulated market was initialized with equal reserves of YES and NO tokens (1000 units each), implying a 50\% initial probability for each outcome.
    
    \item \textbf{Information Scenarios}: We created three information scenarios:
    \begin{itemize}
        \item Scenario A: Gradual information revelation favoring the YES outcome through sequential trades
        \item Scenario B: Oscillating information with contradictory signals but a final bias toward YES
        \item Scenario C: Sharp information shock strongly favoring YES
    \end{itemize}
    
    \item \textbf{Trading Sequences}: For each information scenario, we implemented two distinct trading sequences that reflected identical net information but differed in operation order:
    \begin{itemize}
        \item Sequence 1: Liquidity additions primarily before significant informed trades
        \item Sequence 2: Liquidity additions primarily after significant informed trades
    \end{itemize}
    
    \item \textbf{Path Manipulation Factors}: We systematically varied:
    \begin{itemize}
        \item Liquidity addition timing (early vs. late)
        \item Liquidity addition size (measured by $\alpha$, from 10\% to 100\%)
        \item Trade concentration (single large trade vs. multiple smaller trades)
    \end{itemize}
    
    \item \textbf{Measurement}: We tracked the final implied probability (price of YES token) across different path variations and calculated the path-induced price divergence.
\end{enumerate}

\begin{table}[h]
\centering
\caption{Operation Sequences for Simulation Scenarios}
\begin{tabular}{llll}
\toprule
\textbf{Scenario} & \textbf{Sequence} & \textbf{Operations} & \textbf{P(Yes)} \\
\midrule
Gradual & Path 1 & (+) 100 NO $\rightarrow$ (+) 20\% liquidity $\rightarrow$ & \\
        &        & (+) 150 NO $\rightarrow$ (+) 200 NO & 61.3\% \\
        & Path 2 & (+) 100 NO $\rightarrow$ (+) 150 NO $\rightarrow$ & \\
        &        & (+) 200 NO $\rightarrow$ (+) 20\% liquidity & 62.3\% \\
\midrule
Oscillating & Path 1 & (+) 200 NO $\rightarrow$ (+) 100 YES $\rightarrow$ & \\
            &        & (+) 30\% liquidity $\rightarrow$ (+) 150 NO & 58.1\% \\
            & Path 2 & (+) 200 NO $\rightarrow$ (+) 100 YES $\rightarrow$ & \\
            &        & (+) 150 NO $\rightarrow$ (+) 30\% liquidity & 58.8\% \\
\midrule
Sharp Shock & Path 1 & (+) 50\% liquidity $\rightarrow$ (+) 500 NO & 63.4\% \\
            & Path 2 & (+) 500 NO $\rightarrow$ (+) 50\% liquidity & 69.2\% \\
\bottomrule
\end{tabular}
\label{tab:operation_sequences}
\end{table}

\subsubsection{Simulation Results}

Our simulation revealed significant divergence in final prices despite identical net information flow across all three scenarios. 

The key findings include:

\begin{enumerate}
    \item \textbf{Significant Price Divergence}: Across all scenarios, we observed divergence in final prices despite identical net information. The magnitude of divergence ranged from 1.0 to 10.8 percentage points in implied probability.
    
    \item \textbf{Liquidity Timing Effects}: Markets where large liquidity additions occurred before significant informed trading (Path 1) showed systematically different final probabilities compared to markets where liquidity was added after trading (Path 2). The direction and magnitude of this effect varied based on the information scenario.
    
    \item \textbf{Sharp Information Sensitivity}: The "sharp information shock" scenario exhibited the largest path-dependent effects, with a probability divergence of 5.8 percentage points between Path 1 (63.4\%) and Path 2 (69.2\%) when using a 50\% liquidity addition ($\alpha = 0.5$).
    
    \item \textbf{Liquidity Size Impact}: We found that the magnitude of path dependence scaled with the size of liquidity additions. In extended simulations with extreme parameters (100\% liquidity addition, $\alpha = 1.0$), the probability divergence in the shock scenario reached 10.8 percentage points.
    
    \item \textbf{Trade Fragmentation Effects}: Markets with fragmented trading (multiple small trades) showed less path dependence than those with concentrated trading (few large trades), but the effect remained significant at 1.0-2.5 percentage points.
\end{enumerate}

Table \ref{tab:divergence_by_alpha} shows how the probability divergence scales with the liquidity addition size ($\alpha$) across different scenarios:

\begin{table}[h]
\centering
\caption{Probability Divergence by Liquidity Size ($\alpha$)}
\begin{tabular}{lccc}
\toprule
\textbf{Liquidity Size} & \textbf{Gradual} & \textbf{Oscillating} & \textbf{Sharp Shock} \\
\textbf{($\alpha$)} & \textbf{Scenario} & \textbf{Scenario} & \textbf{Scenario} \\
\midrule
10\% & 0.5 pp & 0.4 pp & 3.2 pp \\
20\% & 1.0 pp & 0.7 pp & 5.8 pp \\
50\% & 2.2 pp & 1.5 pp & 5.8 pp \\
75\% & 3.1 pp & 2.2 pp & 8.3 pp \\
100\% & 3.9 pp & 2.8 pp & 10.8 pp \\
\bottomrule
\end{tabular}
\label{tab:divergence_by_alpha}
\end{table}

To validate the statistical significance of these findings, we conducted Monte Carlo simulations with 1,000 randomized path variations for each scenario. The results confirmed that the observed divergences were not due to chance ($p < 0.001$ for all scenarios).

\subsubsection{Mathematical Explanation of Divergence}

The observed divergence can be explained by examining the constant product formula under different operation sequences. For a market with reserves $x$ (YES tokens) and $y$ (NO tokens):

When liquidity is added before a trade (Path 1):
\begin{align}
    x_1 &= x_0 \times (1 + \alpha) \\
    y_1 &= y_0 \times (1 + \alpha) \\
    k_1 &= k_0 \times (1 + \alpha)^2
\end{align}

After a subsequent trade adding $\Delta y$ NO tokens:
\begin{align}
    y_2 &= y_1 + \Delta y = y_0(1 + \alpha) + \Delta y \\
    x_2 &= \frac{k_1}{y_2} = \frac{k_0(1 + \alpha)^2}{y_0(1 + \alpha) + \Delta y}
\end{align}

The final implied probability is:
\begin{align}
    P_{YES,1} = \frac{y_2}{x_2 + y_2} = \frac{y_0(1 + \alpha) + \Delta y}{\frac{k_0(1 + \alpha)^2}{y_0(1 + \alpha) + \Delta y} + y_0(1 + \alpha) + \Delta y}
\end{align}

In contrast, when the trade occurs before liquidity addition (Path 2):
\begin{align}
    y_1' &= y_0 + \Delta y \\
    x_1' &= \frac{k_0}{y_1'} = \frac{k_0}{y_0 + \Delta y}
\end{align}

After the subsequent liquidity addition:
\begin{align}
    y_2' &= y_1' \times (1 + \alpha) = (y_0 + \Delta y)(1 + \alpha) \\
    x_2' &= x_1' \times (1 + \alpha) = \frac{k_0(1 + \alpha)}{y_0 + \Delta y}
\end{align}

The final implied probability is:
\begin{align}
    P_{YES,2} = \frac{y_2'}{x_2' + y_2'} = \frac{(y_0 + \Delta y)(1 + \alpha)}{\frac{k_0(1 + \alpha)}{y_0 + \Delta y} + (y_0 + \Delta y)(1 + \alpha)}
\end{align}

Algebraic manipulation confirms that $P_{YES,1} \neq P_{YES,2}$ for all $\alpha > 0$ and $\Delta y \neq 0$, with the difference increasing monotonically with both $\alpha$ and $\Delta y$.

\subsubsection{Implications for Prediction Markets}

These simulation results provide compelling evidence that path dependence is not merely a theoretical concern but can materially impact the prices in AMM-based prediction markets. For context, in prediction markets, a difference of 5-10 percentage points in probability estimation is generally considered substantial, as it can significantly impact decision-making based on these forecasts.

The divergences we observed are particularly notable because:

\begin{enumerate}
    \item They occurred despite identical net information (same total trades and liquidity)
    \item They persisted regardless of market conditions (gradual, oscillating, or shock information)
    \item They scaled predictably with liquidity size and trade concentration
    \item They reached economically significant magnitudes even with realistic parameter values
\end{enumerate}

These findings reinforce our central thesis that AMM-based prediction markets cannot function as pure information aggregators due to their inherent path dependence. The price signals in these markets necessarily encode both informational content and artifacts of the specific historical path of operations.

\section{Implications}
\label{sec:implications}

\subsection{For Market Efficiency Theory}
Our findings challenge traditional conceptualizations of market efficiency. While the Efficient Market Hypothesis doesn't explicitly require path independence, the demonstrated path dependence in AMMs suggests that prices encode historical information beyond current market beliefs.

\subsection{For Mechanism Design}
Designers of prediction markets and other AMM-based systems should be aware that their mechanisms have inherent path dependence. This property may need to be accounted for when interpreting market signals or designing governance mechanisms.

\subsection{For Price Interpretation}
Users of AMM-based prediction markets should understand that prices reflect not just information but also the specific history of that market. Identical markets initialized differently or experiencing different trade sequences may show different prices despite identical information.

\subsection{For DeFi Protocol Development}
Based on our findings regarding path dependence in AMM-based markets, we offer the following recommendations for DeFi protocol developers:

\begin{enumerate}
    \item \textbf{Explicit Acknowledgment of Path Dependence}: Protocol documentation should explicitly acknowledge the path-dependent nature of AMM designs, particularly when these mechanisms are used for price discovery or oracle functions. Users and developers should understand that the current state reflects not just market information but also the specific history of operations.
    
    \item \textbf{Path-Aware Initialization Methods}: Develop initialization procedures for AMM pools that mitigate initial path dependence effects. One approach is to implement ``virtual trading'' during pool initialization to simulate diverse transaction paths before the pool becomes publicly accessible.
    
    \item \textbf{Meta-Pool Architectural Designs}: Consider meta-pool architectures that aggregate multiple independent AMM pools tracking the same assets. By averaging across multiple pools with different histories, the aggregate price signal may be less affected by path-specific artifacts.
    
    \item \textbf{Path Dependence Indicators}: Implement on-chain metrics that quantify the potential path dependence effect in a given pool. For example, a ``path sensitivity index'' could be calculated based on recent liquidity-to-swap ratios and displayed to users, indicating how susceptible the current price might be to path effects.
    
    \item \textbf{Bounded-Path Designs}: Explore alternative AMM designs that bound path dependence effects. For instance, mechanisms that periodically ``reset'' the pool to reflect only current inventory ratios, or designs that incorporate time-weighted average prices rather than instantaneous ratios.
    
    \item \textbf{Economic Mitigations}: Implement economic incentives that encourage path-neutralizing behavior. For example, fee structures could be adjusted to reward trades and liquidity additions that reduce path-dependent price distortions.
    
    \item \textbf{Hybrid Oracle Solutions}: For applications requiring high-fidelity price signals (such as lending protocols), combine AMM price feeds with other oracle mechanisms that have different properties. By aggregating signals from multiple source types, the impact of path dependence can be reduced.
    
    \item \textbf{Simulation Tools for Developers}: Create simulation frameworks that allow protocol developers to test their AMM designs under various operation sequences to quantify path dependence effects before deployment.
    
    \item \textbf{Path-Aware Governance Mechanisms}: Design governance mechanisms that can intervene when path dependence leads to significant price distortions, particularly in markets used for critical financial infrastructure.
    
    \item \textbf{Transparency Requirements}: For prediction markets specifically, implement mandatory disclosures about the potential impact of path dependence on price interpretation, helping users properly contextualize the probability estimates derived from AMM prices.
\end{enumerate}

These recommendations aim to help protocol developers account for the inherent path dependence of AMM systems, either by mitigating its effects or by providing users with the necessary context to properly interpret prices. While path dependence cannot be eliminated from constant product AMMs without fundamentally altering their design, its impacts can be managed through thoughtful protocol engineering and transparent communication.

\section{Conclusion}
\label{sec:conclusion}

This paper has demonstrated the mathematical certainty of path dependence in AMM-based markets. Through formal proofs and empirical evidence, we've shown that the sequence of operations in an AMM determines its final state, challenging the notion that market prices purely reflect information.

This finding has particular relevance for decentralized prediction markets, which often claim to function as "truth machines." Our analysis shows that such claims must be qualified—these markets reflect not just information but also their specific history.

Future research could explore the extent to which traditional markets also exhibit path dependence, whether alternative AMM designs could mitigate path dependence, and how to account for this property when interpreting market signals.

By understanding the mathematical properties of market mechanisms, we gain insights not just into these specific systems but potentially into the nature of markets in general.

\bibliographystyle{plainnat}
\bibliography{references}

\end{document}